\documentclass[HYPER]{JHEP}
\usepackage{graphicx,amssymb,bm,latexsym,amsmath,mathtools}
\usepackage{epstopdf}
\usepackage{caption}
\usepackage{subcaption}
\usepackage[section]{placeins}
\usepackage{float}
\usepackage{capt-of}
\usepackage{tabu}
\usepackage[toc,page]{appendix}
\usepackage[toc,page]{appendix}
\title{Finite size effect from classical strings in $AdS_3 \times S^3$ with NS-NS flux}
\author{Sorna Prava Barik and Kamal L. Panigrahi \\
	Department of Physics,\\Indian Institute of Technology Kharagpur,\\
	Kharagpur-721 302, India\\
	Email: \email{sorna,panigrahi@phy.iitkgp.ac.in}}

\abstract{We study the finite size effect of rigidly rotating strings and closed folded strings in $AdS_3\times S^3$ geometry with NS-NS B-field. We calculate the classical exponential corrections to the dispersion relation of infinite size giant magnon and single spike in terms of Lambert $\mathbf{W}-$function. We also write the analytic expression for the dispersion relation of finite size Gubser-Klebanov-Polyakov (GKP) string in the form of Lambert $\mathbf{W}-$function.}
\begin{document}
	\section{Introduction}
AdS/CFT duality \cite{Maldacena:1997re,Witten:1998qj,Gubser:1998bc} relates $\mathcal{N}=4$ super Yang-Mills (SYM) theory with type IIB superstring theory in $AdS_5 \times S^5$ geometry and vice versa. The traditional way of checking the correspondence is to find the full spectrum on both sides of the duality and compare the scaling dimensions of local gauge-invariant operators in planar limit in the gauge theory side with the energies of free string states in the string theory side. The planar integrability of gauge theory as well as string theory has played a vital role in understanding the duality conjecture better. In this context, it was first observed by Minahan and Zarembo that the one-loop dilatation operators of the SU(2) sector of $\mathcal{N}=4$ SYM theory can be identified with the Hamiltonian of Heigenberg spin chain \cite{Minahan:2002ve,Beisert:2003jj,Beisert:2003yb}. Spin chain Hamiltonian can be diagonalized using Bethe ansatz. Later on the all-loop asymptotic Bethe ansatz was proposed in  \cite{Beisert:2004hm} to compute the higher loop dilatation operators. The computation predicts conformal dimension for one magnon state upto all order in a spin chain of length $J+1$ without having any wrapping correction term (when the size of the spin chain is infinite i.e $J\rightarrow\infty$) 
	\begin{eqnarray}
	\Delta=J+\sqrt{1+\frac{\lambda}{\pi^2}\sin^2\frac{p}{2}},~~~~~\lambda=g_{\text{YM}}^2N,
	\end{eqnarray} 
where $p$ is the magnon momentum. Over the last few years, a large class of semiclassical strings (i.e rotating, spinning, pulsating etc.) has been studied in $AdS_5 \times S^5$ geometry to find the string states dual of the gauge theory operators, see e.g. \cite{Gubser:2002tv,Dimov:2004xi,Kruczenski:2004wg,Lee:2008sk,Barik:2018haz,Barik:2017opb}. Semiclassical solitonic solutions with large quantum numbers represent the highly excited string states and correspond to the gauge invariant operators with small anomalous dimensions. The string theory dual of spin chain magnons are giant magnons, which are open strings rotating in $R\times S^2$ geometry with large spin \cite{Hofman:2006xt}\footnote[1]{giant magnons are long strings with the limit $J\rightarrow\infty, \lambda = {\rm fixed}, p = {\rm fixed}, E-J = {\rm finite}$.}. 
	The dispersion relation for the open string rotating on a 2-sphere with angular extent $\Delta \phi $ is given as
	\begin{eqnarray}
	E-J=\frac{\sqrt{\lambda}}{\pi} \sin\frac{\Delta \phi}{2}.
	\end{eqnarray}
	The mapping of gauge operators to the string states relate the magnon momentum $p$ to the geometrical angle $\Delta\phi$ at large 't Hooft coupling limit ($\lambda\rightarrow\infty$).
	Giant magnons, the elementary string excitations can be used to build closed strings and multi-soliton solutions. The Gubser-Klebanov-Polyakov (GKP) closed folded string can be formed by superimposing two giant magnons with maximum angular extent $\Delta\phi=\pi$ with following dispersion relation
	\begin{eqnarray}
	E-J=\frac{2\sqrt{\lambda}}{\pi}. 
	\end{eqnarray}
	
		The asymptotic Bethe ansatz is a powerful technique to compute the anomalous dimensions only when the length of the spin chain is infinite or much larger than the loop order. Beyond the critical loop order, the range of the spin chain interactions become more significant than the length of the spin chain. In this case, wrapping corrections have to be taken into account to get correct conformal dimensions of operators \cite{Sieg:2005kd,SchaferNameki:2005tn,SchaferNameki:2006ey,Ambjorn:2005wa,Kotikov:2007cy,Gromov:2009tv,Gromov:2013pga}. In CFT, the wrapping correction can be considered as the higher genus correction to the dilatation operator while in string theory side wrapping effect is due to the finite circumference of the cylindrical worldsheet. On both sides, wrapping effect can be neglected by considering infinite global charges. The anomalous dimension with wrapping corrections can be related to the dispersion relation with finite conserved charges in string theory side.  In this context, finite size corrections to the giant magnons has been studied extensively \cite{Arutyunov:2006gs,Astolfi:2007uz,Klose:2008rx,Ramadanovic:2008qd,Minahan:2008re,Shenderovich:2008bs,Ahn:2008gd}. For related work see \cite{Jain:2008mt,Ahn:2008sk,Bykov:2008bj,Grignani:2008te,Bozhilov:2010rf,Ahn:2010da,Ahn:2014aqa,Ahn:2016egk,Panigrahi:2018xuv}. Both wrapping corrections to the anomalous dimensions and finite size corrections to the scaling relations have the form of exponential corrections with respect to infinite solutions. Recently, the classical exponential corrections to the dispersion relations for GKP strings, giant magnons and single spikes have been expressed in terms of Lambert $\mathbf{W}-$function \cite{Floratos:2013cia,Floratos:2014gqa,Linardopoulos:2015hxa}.
	
	Integrability of superstring theory on $AdS_3\times S^3 \times T^4$ geometry supported by a combination of the NS-NS and RR fluxes has  been studied and tree-level S-matrix has been constructed for generic value of $q$ \cite{Hoare:2013pma,Hoare:2013ida,OhlssonSax:2018hgc}. More recently, all-loop S-matrix of type IIB string theory on $AdS_3\times S^3 \times T^4$ and $AdS_3\times S^3 \times S^3 \times S^1$ with mixed flux upto the dressing factors has been proposed using off-shell symmetry algebra \cite{Lloyd:2014bsa,Borsato:2015mma}\footnote[2]{see \cite{Sfondrini:2014via} for a review and references therein to get the detailed approach to this problem.}. It is believed that the pure RR flux model (parameter of the NSNS flux $q=0$) can be described by Bethe ansatz  while the pure NSNS flux model $(q = 1)$ can be described by WZW model. The exact form of the dispersion relation for dyonic giant magnon
	soliton on $AdS_3 \times S^3 \times T^4$ with mixed flux has been constructed in \cite{Hoare:2013lja} and the leading finite size correction for the same has been computed in \cite{Ahn:2014tua}. It is important to learn
	how the NS-NS B-field modifies the anomalous dimension due the wrapping and the finite size correction to the usual scaling relations in various orders the parameter parametrizing the relations in terms of the Lambert $\mathbf{W}-$function. Motivated by the recent surge of interest in the study of semiclassical strings in the context of AdS/CFT duality and integrability in gauge and string theory, in this paper, we study the finite size classical strings in $AdS_3\times S^3$ with NS-NS flux and express the exponential corrections to the dispersion relations in terms of Lambert $\mathbf{W}-$function. \\
	
	The rest of the paper is organized as follows. In section 2, we study finite size giant magnon and spike string in $R \times S^2$ geometry with NS-NS B-field. We compute the  exponential corrections to the infinite size dispersion relations and write the results in terms of Lambert $\mathbf{W}$-function. In section 3, we discuss closed folded strings in $R \times S^2$ NS-NS flux and finite size corrections has been calculated by expanding the charges for large values. By applying Lagrange inversion formalism, we write the scaling relation in terms of Lambert $\mathbf{W}$-function. Section 4 is devoted to the concluding remarks. 
	\section{Rigidly Rotating Strings}
	In this section, we wish to study the finite size giant magnon and single spike in $R\times S^2$ background with NS-NS B-field. The relevant metric and B-field can be written as follows,
	\begin{eqnarray}
	ds^2=-dt^2+d\theta^2+\sin^2\theta d\phi^2, \nonumber \\
	B_{NS}=-B \sin\theta d\theta \wedge d\phi.
	\end{eqnarray}
	The Polyakov action in conformal gauge for the choosen background takes the form
	\begin{eqnarray}
	S=\dfrac{\sqrt{\lambda}}{4 \pi}\int d\tau d\sigma[-(\dot{t}^2-t'^2)+\dot{\theta}^2-\theta'^2+\sin^2\theta(\dot{\phi}^2-\phi'^2)-2B\sin\theta(\dot{\theta}\phi'-\theta'\dot{\phi})],\nonumber
	\end{eqnarray}
	where $\lambda$ is the 't Hooft coupling constant. Dot and prime denotes  the derivative with respect to $\tau$ and $\sigma$ respectively. We start with the following ansatz
	\begin{eqnarray}
	t=t(\tau),~~~ \theta=\theta(y),~~~\phi=\omega\tau+h(y),
	\end{eqnarray}
	where $y=\sigma-\nu\omega\tau$. The equation of motion for $t$ can be solved for $t=k \tau$, where $k$ is the integration constant. Substituting $z=\cos\theta$, we get the following set of  Virasoro constraints
	\begin{eqnarray}\label{eqnh}
	h'=\frac{\nu\omega^2(z^2-z_\omega^2)}{(1-\nu^2\omega^2)(1-z^2)},
	\end{eqnarray}
	\begin{eqnarray}\label{eqnz}
	z'^2 =\frac{\omega^2}{(1-\nu^2 \omega^2)^2}\left(z^2-z_\omega^2 \right) \left( z_\nu^2-z^2\right),
	\end{eqnarray}
	where $z_\omega^2=1-\frac{k^2}{\omega^2}$,\, $ z_\nu^2=1-k^2\nu^2$ and the prime denotes the derivative with respect to $y$. Now the conserved charges and the deficit angle become
	\begin{eqnarray}
	E&=&\frac{\sqrt{\lambda}k}{\pi} \frac{(1-\nu^2 \omega^2)}{\omega} \int_{z_{min}}^{z_{max}} dz \frac{1}{\sqrt{(z^2-z_\omega^2)(z_\nu^2-z^2)}}, \\
	J&=&\frac{\sqrt{\lambda}}{\pi}  \int_{z_{min}}^{z_{max}} dz \left[\frac{(z_\nu^2-z^2)}{\sqrt{(z^2-z_\omega^2)(z_\nu^2-z^2)}}+B\right] , \\
	\Delta \phi &=&2\int_{z_{min}}^{z_{max}} dz \frac{h'}{z'}=2\nu\omega\int_{z_{min}}^{z_{max}} dz \frac{z^2-z_\omega^2}{(1-z^2)\sqrt{(z^2-z_\omega^2)(z_\nu^2-z^2)}}.
	\end{eqnarray}
	We have following set of string solutions depending upon the value of $\omega$ and $\nu$:
	\begin{eqnarray}
	\nu \omega &\le& 1 ~~~\text{where}~ \omega \ge 1 ~ \text{and}\,\nu \le 1 ~~~~~~\text{Giant Magnon solution} \nonumber \\ 
	\nu \omega &\ge& 1 ~~~\text{where}~ \omega \ge 1 ~ \text{and}\,\nu \le 1 ~~~~~~\text{Single Spike  solution}
	\end{eqnarray}
	\subsection{Giant Magnon Solution: $0 \le k\nu \le \frac{k}{\omega}\le 1$}
	In this case we have,
	\begin{eqnarray}
	z_{min}^2=z_\omega^2=1-\frac{k^2}{\omega^2}~~\text{and}~~z_{max}^2=z_\nu^2=1-k^2\nu^2.
	\end{eqnarray}
	The Virasoro constraint equations (\ref{eqnh}) and (\ref{eqnz}) can be solved for 
	\begin{eqnarray}
	z(\tau,\sigma)&=&\cos\theta=\sqrt{1-k^2\nu^2}\,\text{dn}\left(\frac{\omega \sqrt{1-k^2\nu^2} }{1-\nu^2 \omega^2}(\sigma-\nu\omega\tau),1-\epsilon^2\right),\\
	h(z)&=&\frac{1}{\sqrt{1-k^2\nu^2}}\left[\frac{1}{\nu\omega}\boldmath{\Pi}\left(1-\frac{1}{\nu^2\omega^2},\arcsin\left(\frac{1}{\sqrt{1-\epsilon^2}}\sqrt{1-\frac{z^2}{1-k^2\nu^2}}\right),1-\epsilon^2\right)-\right. \nonumber\\ 
	&-&\left.\nu\omega \boldmath{F}\left(\arcsin\left(\frac{1}{\sqrt{1-\epsilon^2}}\sqrt{1-\frac{z^2}{1-k^2\nu^2}}\right),1-\epsilon^2\right)\right],
 	\end{eqnarray}
 	where, we have defined $	\epsilon^2=\frac{z^2_{min}}{z^2_{max}}$.\\
	The magnon momentum which is defined as the angular difference between the two end points of the string on $S^2$ is given by
	\begin{eqnarray}\label{gmp}
	p=\Delta \phi=\frac{\pi }{\boldmath{K}(\epsilon^2)}\boldmath{F}(\arcsin z_{max},\epsilon^2)+\frac{2z_{max}(1-\epsilon^2)\boldmath{K}(1-\epsilon^2)}{\sqrt{1-z_{max}^2}\sqrt{1-\epsilon^2z_{max}^2}\boldmath{K}(\epsilon^2)} \nonumber\\
	\times\left(\boldmath{K}(\epsilon^2)-\boldmath{\Pi}\left(\frac{(1-z_{max}^2)\epsilon^2}{1-\epsilon^2z_{max}^2},\epsilon^2\right)\right).
	\end{eqnarray}
	The conserved energy and angular momentum can be written as
	\begin{eqnarray}\label{gmcharge}
	\mathcal{E}&=&\frac{\pi}{\sqrt{\lambda}}E=\frac{z_{max}(1-\epsilon^2)}{\sqrt{1-\epsilon^2z_{max}^2}}\boldmath{K}(1-\epsilon^2),\nonumber\\ 
	\mathcal{J}&=&\frac{\pi}{\sqrt{\lambda}}J=z_{max}[\boldmath{K}(1-\epsilon^2)-\boldmath{E}(1-\epsilon^2)+B(1-\epsilon)],
	\end{eqnarray}
	and 
	\begin{eqnarray}
	\mathcal{E}-\mathcal{J}=z_{max}\left[\boldmath{E}(1-\epsilon^2)-\left(1-\frac{1-\epsilon^2}{\sqrt{1-\epsilon^2z^2_{max}}}\right)\boldmath{K}(1-\epsilon^2)-B(1-\epsilon)\right].
	\end{eqnarray}
	In the limit, $\frac{k}{\omega}=1$ and $k\nu\le1$, we have infinite energy and angular momentum where as their difference is finite. But we are interested in the regime where the conserved charges are large but finite, which can be obtained by taking limit $\epsilon \rightarrow 0$. But in this limit the conserved charges have logarithmic singularities due to presence of following elliptic functions
	\begin{eqnarray}
	\boldmath{K}(1-\epsilon^2)=\sum_{n=0}^{\infty} \epsilon^{2n}~ (2d_n \ln\epsilon+h_n), \\
	\boldmath{K}(1-\epsilon^2)-\boldmath{E}(1-\epsilon^2)=\sum_{n=0}^{\infty} \epsilon^{2n}~ (2c_n \ln\epsilon+b_n).
	\end{eqnarray}
	Here the coefficients of the above series are given by
	\begin{eqnarray}
	d_n=-\frac{1}{2}\left(\frac{(2n-1)!!}{(2n)!!}\right)^2,~~~h_n=-4d_n(\ln2+H_n-H_{2n}),\nonumber\\
	c_n=-\frac{d_n}{2n-1},~~~b_n=-4c_n\left(\ln2+H_n-H_{2n}+\frac{1}{2(2n-1)}\right),\nonumber
	\end{eqnarray}
	where~~ $H_n=\sum_{k=1}^{n}\frac{1}{k}$.\\
	The expression of magnon momentum after eliminating logarithms 
	\begin{eqnarray}
	p=\frac{\pi }{\boldmath{K}(\epsilon^2)}\boldmath{F}( a,\epsilon^2)+\frac{2(1-\epsilon^2)\tan a}{\boldmath{K}(\epsilon^2)\sqrt{1-\epsilon^2\sin^2 a}} 
	\left(\boldmath{K}(\epsilon^2)-\boldmath{\Pi}\left(\frac{\epsilon^2 \cos^2a}{1-\epsilon^2\sin^2a},\epsilon^2\right)\right) \nonumber \\
	\times\left[\sum_{n=0}^{\infty} h_n \epsilon^{2n}+\frac{\sum_{n=0}^{\infty} d_n \epsilon^{2n}}{\sum_{n=0}^{\infty} c_n \epsilon^{2n}}\left(\mathcal{J} \csc a-B (1-\epsilon)-\sum_{n=0}^{\infty} b_n \epsilon^{2n} \right)\right],\nonumber \\
	\end{eqnarray}
	where we have substituted $z_{max}=\sin a$. From equations (\ref{gmp}) and (\ref{gmcharge}), we can see the conserved charges and magnon momentum are function of two parameter i.e $z_{max}$ and $\epsilon$. So to obtain the dispersion relation $\mathcal{E}=\mathcal{E} (p,\mathcal{J})$, first we need to express $z_{max}=z_{max}(\epsilon,p,\mathcal{J})$ and $\epsilon=\epsilon(p,\mathcal{J})$. In the process to get $z_{max}$, we have expanded $p$ around both $\epsilon=0$ and $a=\frac{p}{2}$. Then the series is inverted to find $a$ and from there $z_{max}(\epsilon,p,\mathcal{J})$. The result is given in appendix A, which has been computed by using mathematica. 
	From the expression of angular momentum 
	\begin{eqnarray}\label{lne}
	\ln \epsilon=\frac{1}{2}\left[\frac{\frac{\mathcal{J}}{z_{max}(\epsilon,p,\mathcal{J})}-B(1-\epsilon)-b_0}{c_0}-\sum_{n=1}^{\infty}\frac{b_n}{c_0} \epsilon^{2n}\right]\sum_{k=0}^{\infty}\left(-\sum_{n=1}^{\infty} \frac{c_n}{c_0} \epsilon^{2n}\right)^k.
	\end{eqnarray}
	The above equation can be written in a simpler form as following 
	\begin{eqnarray}\label{e}
	\epsilon_0=\epsilon~\exp\left[-\left(\sum_{n=1}^{\infty} \text{a}_n(p,\mathcal{J}) \epsilon^n\right)\right],
	\end{eqnarray}
	where the coefficients $\text{a}_n$ can be determined from (\ref{lne}) by substituting the value of $z_{max}(\epsilon,p,\mathcal{J})$  and 
	\begin{eqnarray}
	\epsilon_0=4~ \exp\left[-\mathcal{J}\csc \frac{p}{2}-1+B\right]
	\end{eqnarray}
	is the zero order term in the series expansion of (\ref{lne}) around $\epsilon$. Now we invert the series (\ref{e}) via Lagrange-B$\ddot{\text{u}}$rmann formula and end up with the result $\epsilon=\epsilon(p,\mathcal{J})$, in a series involving Bell polynomial,
	\begin{eqnarray}
	\epsilon=\sum_{n=1}^{\infty} \frac{\epsilon_0^n}{n!} \sum_{k\le n-1} n^k \mathcal{B}_{n-1,k}
	\end{eqnarray}
	where $\mathcal{B}_{n-1,k}=\mathcal{B}_{n-1,k}(1!\text{a}_1,2!\text{a}_2,3!\text{a}_3......(n-k)!\text{a}_{n-k})$ is the Bell polynomial and the leading order term in $\epsilon^2$ is found to be
	\begin{eqnarray}
	\epsilon^2_{\text{leading}}=-\frac{1}{\mathcal{J}^2}\tan^2\frac{p}{2} ~\mathbf{W}.
	\end{eqnarray}
	Here $\mathbf{W}=\mathbf{W}\left(-16\mathcal{J}^2 \cot^2\frac{p}{2}\, \exp\left[-2-2\mathcal{J}\csc\frac{p}{2}+2B\right]\right)$ defines the principal branch of Lambert $\mathbf{W}$-function with argument $(-16\mathcal{J}^2 \cot^2\frac{p}{2}\, \exp[-2-2\mathcal{J}\csc\frac{p}{2}+2B])$.\\
	The finite size correction to the giant magnon dispersion relation can be written in terms of Lambert $\mathbf{W}$-function by plugging this result into expression for $(\mathcal{E}-\mathcal{J})$ as
	\begin{eqnarray}
	\mathcal{E}-\mathcal{J}&=&(1-B)\sin\frac{p}{2}+\frac{B}{4\mathcal{J}}\left[4\tan\frac{p}{2}\sin\frac{p}{2}\sqrt{-\mathbf{W}}+\tan^2\frac{p}{2}(1+\cos p)\mathbf{W}\right]+ \nonumber \\
	&+&\frac{1}{4\mathcal{J}^2}\tan^2\frac{p}{2} \sin^3\frac{p}{2}\left[\mathbf{W}+\frac{\mathbf{W}^2}{2}+3B\left[\cot^2\frac{p}{2}\mathbf{W}+\frac{2}{3}\csc\frac{p}{2} \cot\frac{p}{2}(-\mathbf{W})^{3/2}+\frac{\mathbf{W}^2}{2}\right]\right]+...\nonumber\\
	\end{eqnarray}
	In infinite $\mathcal{J}$ limit, this result exactly reduces to the dispersion relation of infinite size giant magnon with NS-NS flux \cite{Jain:2008mt} and turning the B-field off, it is in complete agreement with \cite{Floratos:2014gqa}.
	\subsection{Single Spike Solution: $0  \le \frac{k}{\omega}  \le k\nu \le 1 $}
	This is the case where,
	\begin{eqnarray}
	z_{max}^2=1-\frac{k^2}{\omega^2}~~\text{and}~~ z_{min}^2=1-k^2\nu^2. 
	\end{eqnarray}
	With this choice, substituting $\epsilon^2=\frac{z^2_{min}}{z^2_{max}}$, the solutions of  Virasoro constraint equations (\ref{eqnh}) and (\ref{eqnz}) becomes 
	\begin{eqnarray}
	z(\tau,\sigma)&=&\cos\theta=\sqrt{1-\frac{k^2}{\omega^2}}\,\text{dn}\left(\frac{\omega \sqrt{1-\frac{k^2}{\omega^2}} }{1-\nu^2 \omega^2}(\sigma-\nu\omega\tau),1-\epsilon^2\right),\\
	h(z)&=&\frac{\nu\omega}{\sqrt{1-\frac{k^2}{\omega^2}}}\left[\boldmath{\Pi}\left(1-\nu^2\omega^2,\arcsin\left(\frac{1}{\sqrt{1-\epsilon^2}}\sqrt{1-\frac{z^2}{1-\frac{k^2}{\omega^2}}}\right),1-\epsilon^2\right)-\right. \nonumber\\ 
	&-&\left. \boldmath{F}\left(\arcsin\left(\frac{1}{\sqrt{1-\epsilon^2}}\sqrt{1-\frac{z^2}{1-\frac{k^2}{\omega^2}}}\right),1-\epsilon^2\right)\right].
	\end{eqnarray}
	In case of single spike solution for infinite size limit (i.e $k\nu=1$ and $\frac{k}{\omega}\le 1$), the energy and angular deficit become infinite while their difference is finite with finite angular momentum. Using the same technique as in case of giant magnon, we proceed to find the dispersion relation for finite size spike solution. The deficit angle become
	\begin{eqnarray}
	p&=&\frac{2z_{max}(1-\epsilon^2)\boldmath{K}(1-\epsilon^2)}{\sqrt{1-z_{max}^2}\sqrt{1-\epsilon^2z_{max}^2}\boldmath{K}(\epsilon^2)} 
	\boldmath{\Pi}\left(\frac{(1-z_{max}^2)\epsilon^2}{1-\epsilon^2z_{max}^2},\epsilon^2\right)-\frac{\pi }{\boldmath{K}(\epsilon^2)}\boldmath{F}(\arcsin z_{max},\epsilon^2).\nonumber\\
	\end{eqnarray} 
	The conserved energy and angular momentum turn out to be:
	\begin{eqnarray}
	\mathcal{E}&=&\frac{\pi}{\sqrt{\lambda}} E=\frac{z_{max}(1-\epsilon^2)}{\sqrt{1-z_{max}^2}}\boldmath{K}(1-\epsilon^2),\nonumber\\
	\mathcal{J}&=&\frac{\pi}{\sqrt{\lambda}} J=z_{max}[\boldmath{E}(1-\epsilon^2)-\epsilon^2 \boldmath{K}(1-\epsilon^2)-B(1-\epsilon)],\nonumber
	\end{eqnarray}
	leading to the following expression for $\left(\mathcal{E}-\frac{p}{2}\right)$
	\begin{eqnarray}\label{spikydr}
	\mathcal{E}-\frac{p}{2}=\frac{z_{max}(1-\epsilon^2)}{\sqrt{1-z_{max}^2}}\boldmath{K}(1-\epsilon^2)\left[1-\frac{1}{\sqrt{1-\epsilon^2z_{max}^2}\boldmath{K}(\epsilon^2)}\boldmath{\Pi}\left(\frac{(1-z_{max}^2)\epsilon^2}{1-\epsilon^2z_{max}^2},\epsilon^2\right)\right]+\nonumber\\ +\frac{\pi}{2\boldmath{K}(\epsilon^2)}\boldmath{F}(\arcsin z_{max},\epsilon^2).\nonumber\\
	\end{eqnarray}
	The expression of conserved charges contain logarithmic singularities in the limit $\epsilon \rightarrow 0$ due to presence of elliptic functions $\boldmath{E}(1-\epsilon^2)$ and $\boldmath{K}(1-\epsilon^2)$.
	So we write the angular momentum after eliminating logarithms as
	\begin{eqnarray}
	\mathcal{J}=z_{max}\left[\frac{\sqrt{1-z_{max}^2}\sqrt{1-\epsilon^2z_{max}^2}\boldmath{K}(\epsilon^2)}{2z_{max} \boldmath{\Pi}\left(\frac{(1-z_{max}^2)\epsilon^2}{1-\epsilon^2z_{max}^2},\epsilon^2\right)}\left(p+\frac{\pi}{\boldmath{K}(\epsilon^2)}\boldmath{F}(\arcsin z_{max},\epsilon^2)\right) \times\right. \nonumber \\ 
\times\left. \left(1-\frac{1}{1-\epsilon^2}\frac{\sum_{n=0}^{\infty} c_n \epsilon^{2n}}{\sum_{n=0}^{\infty} d_n \epsilon^{2n}}\right)+\frac{\sum_{n=0}^{\infty} h_n \epsilon^{2n}}{\sum_{n=0}^{\infty} d_n \epsilon^{2n}}\sum_{n=0}^{\infty} c_n \epsilon^{2n}-\sum_{n=0}^{\infty} b_n \epsilon^{2n}-B(1-\epsilon)\right].\nonumber\\
	\end{eqnarray}
	Performing the series expansion of $\mathcal{J}$ expression and then inverting it, we can get $z_{max}(\epsilon,p,\mathcal{J})$.
	From the expression of linear momentum
	\begin{eqnarray}\label{lne1}
	\ln \epsilon=\frac{1}{2}\left[\frac{\sqrt{1-z_{max}^2}\sqrt{1-\epsilon^2z_{max}^2}\boldmath{K}(\epsilon^2)}{2z_{max}d_0(1-\epsilon^2) \boldmath{\Pi}\left(\frac{(1-z_{max}^2)\epsilon^2}{1-\epsilon^2z_{max}^2},\epsilon^2\right)}\left(p+\frac{\pi}{\boldmath{K}(\epsilon^2)}\boldmath{F}(\arcsin z_{max},\epsilon^2)\right)-\sum_{n=0}^{\infty}\frac{h_n}{d_0} \epsilon^{2n}\right]\nonumber\\
	\times \sum_{k=0}^{\infty}\left(-\sum_{n=1}^{\infty} \frac{d_n}{d_0} \epsilon^{2n}\right)^k\nonumber\\
	\end{eqnarray}
	The expression (\ref{lne1}) can be equivalently written in more compact form as
	\begin{eqnarray}\label{e1}
	\epsilon_0=\epsilon~\exp\left[-\left(\sum_{n=1}^{\infty} \text{a}_n(p,\mathcal{J}) \epsilon^n\right)\right],
	\end{eqnarray}
	where the coefficients $\text{a}_n(p,\mathcal{J})$ can be determined from (\ref{lne1}) after putting $z_{max}(\epsilon,p,\mathcal{J})$ to the expression and doing the product expansion around $\epsilon=0$. Here $\epsilon_0$ is the lowest order term from equation(\ref{lne1}) and substituting $\mathcal{J}=\sin\frac{q}{2}$, the expression for $\epsilon_0$ becomes
	\begin{eqnarray}
	\epsilon_0=4~\exp\left[-\frac{1}{2}(p+q)\cot\frac{q}{2}-B\left(1-(p+q)\csc q\right)\right].
	\end{eqnarray}
	Inverting the series (\ref{e1}) by Lagrange-B$\ddot{\text{u}}$rmann inversion formalism, $\epsilon$ can be expressed in terms of Bell polynomials and we write the result for $\epsilon^2$
	upto leading order
	\begin{eqnarray}
	\epsilon^2_{\text{leading}}=-\frac{4}{p^2}\sin^2\frac{q}{2}\, \mathbf{W}\left(-4p^2\csc^2\frac{q}{2}\text{exp}\left[-(p+q)\cot\frac{q}{2}-2B(1-(p+q)\csc q)\right]\right).\nonumber\\
	\end{eqnarray}
	Now inserting $\epsilon(p,q)$ in equation (\ref{spikydr}),  provide the desired result for single spike scaling relation in terms of Lambert $\mathbf{W}$-function
	\begin{eqnarray}
	\mathcal{E}-\frac{p}{2}&=&\frac{q}{2}+B\tan\frac{q}{2}-\frac{B}{p}\sin^2\frac{q}{2}\left[2 \sec\frac{q}{2}\sqrt{-\mathbf{W}}+\mathbf{W}\right]-\frac{1}{p^2}\sin^2\frac{q}{2}\left[\sin^2\frac{q}{2}\tan\frac{q}{2}\left[\frac{\mathbf{W}^2}{2}+\mathbf{W}\right]+\right.\nonumber\\
	&+&\left.B\left(\left(q-\sin q+\left(2+\sec^2\frac{q}{2}\right)\tan\frac{q}{2}\right)\mathbf{W}+2 \sin\frac{q}{2}(-\mathbf{W})^{3/2}+\frac{1}{2}\cos q \tan^3\frac{q}{2}\mathbf{W}^2\right)\right]+...\nonumber\\
	\end{eqnarray}
	The first two terms are the scaling relation of single spike in the infinite volume which can be obtained by substituting $p=\infty$ and putting $B=0$, this result is consistent with the finite size dispersion relation for single spike without having B-field \cite{Floratos:2014gqa}.
	\section{Spinning Folded Strings on $R\times S^2$}
	The ansatz for the spinning string which rotates around the north pole of $S^2\subset S^5$ and correspondingly stretched
	\begin{eqnarray}
	t=k\tau,~~~ \theta=\theta(\sigma),~~~\phi=k\omega\tau.
	\end{eqnarray}
	In this section, we study the closed folded string in $R\times S^2$ geometry with NS-NS B-field
	\begin{eqnarray}
	B_{NS}=-B \sin\theta d\theta \wedge d\phi.
	\end{eqnarray}
	As discussed in the previous section, the spin chain giant magnon momentum can be interpreted in string theory side as a goemetrical angle. Superimposing two giant magnon solution having maximun angular extent ($\Delta\phi=\pi$), GKP solution can be formed\cite{Hofman:2006xt}, which is dual to the spin chain magnon excitations with momentum $p=\pi$.
	From the virasoro constraint 
	\begin{eqnarray}
	\theta'^2=k^2(1-\omega^2\sin^2\theta),
	\end{eqnarray}
	which can be solved for
	\begin{eqnarray}\label{gkpv}
	\sin\theta=\frac{1}{\omega}~\text{sn}[k\omega\sigma,\frac{1}{\omega^2}]~~~~~~\text{for}~\omega^2>1.
	\end{eqnarray}
	From equation (\ref{gkpv}), we can see the string is stretched along $\theta$ direction with $\theta_{max}$
	\begin{eqnarray}
	\sin\theta_{max}=\frac{1}{\omega}.
	\end{eqnarray}
	The coserved energy and angular momentum of the spinning string
	\begin{eqnarray}
	E&=&\frac{\sqrt{\lambda}}{2\pi}\int k d\sigma=\frac{2\sqrt{\lambda}}{\pi} \int_{0}^{\theta_{max}}d\theta \frac{1}{\sqrt{1-\omega^2\sin^2\theta}},\\
	J&=&\frac{\sqrt{\lambda}}{2\pi}\int d\sigma[k\omega \sin^2\theta+B\sin\theta \theta']=\frac{2\sqrt{\lambda}}{\pi} \int_{0}^{\theta_{max}}d\theta \left[\frac{\omega \sin^2\theta}{\sqrt{1-\omega^2\sin^2\theta}}+B \sin\theta\right].\nonumber\\
	\end{eqnarray}
	In case of infinite size limit, the string at north pole approaches to the equator i.e $\theta_{max}=\pi/2$ which imply $\omega=1$. Both energy and angular momentum diverge in this limit while their difference becomes finite
	\begin{eqnarray}\label{infgkp}
	E-J=\frac{2\sqrt{\lambda}}{\pi}(1-B).
	\end{eqnarray}
	Let's proceed to find the dispersion relation for large but finite conserved charges i.e when $\omega^2>1$. We can write the conserved charges in terms of elliptic functions
	\begin{eqnarray}
	E&=&\frac{2\sqrt{\lambda}}{\pi}~ \boldmath{F}(\theta_{max},\omega^2)=\frac{2\sqrt{\lambda}}{\pi\omega}\boldmath{K}\left(\frac{1}{\omega^2}\right),\nonumber \\
	J&=&\frac{2\sqrt{\lambda}}{\pi}~\left[\boldmath{K}\left(\frac{1}{\omega^2}\right)-\boldmath{E}\left(\frac{1}{\omega^2}\right)+B\left(1-\sqrt{1-\frac{1}{\omega^2}}\right)\right].
	\end{eqnarray}
	Now substituting $\epsilon^2=1-\frac{1}{\omega^2}$, we obtain the following set of equations
	\begin{eqnarray}
	\mathcal{E}&=&\frac{\pi}{2\sqrt{\lambda}} E=\sqrt{1-\epsilon^2}\boldmath{K}(1-\epsilon^2), \\
	\mathcal{J}&=&\frac{\pi}{2\sqrt{\lambda}} J=\boldmath{K}(1-\epsilon^2)-\boldmath{E}(1-\epsilon^2)+B(1-\epsilon),\\
	\mathcal{E}-\mathcal{J}&=&(\sqrt{1-\epsilon^2}-1)\boldmath{K}(1-\epsilon^2)+\boldmath{E}(1-\epsilon^2)-B(1-\epsilon).
	\end{eqnarray}
	After isolation of singularity, we can get $\epsilon=\epsilon(\mathcal{J})$ by inverting $\mathcal{J}-$series and can express $\mathcal{E}=\mathcal{E}(\mathcal{J})$ by substituting $\epsilon(\mathcal{J})$ into $\mathcal{E}(\epsilon)$. Angular momentum expression can be solved for $\ln\epsilon$:
	\begin{eqnarray}\label{gmlne}
	\ln\epsilon=\frac{1}{2}\left[\frac{\mathcal{J}-B(1-\epsilon)-b_0}{c_0}-\sum_{n=1}^{\infty}\frac{b_n}{c_0} \epsilon^{2n}\right]\sum_{k=0}^{\infty}\left(-\sum_{n=1}^{\infty} \frac{c_n}{c_0} \epsilon^{2n}\right)^k.
	\end{eqnarray}
	We write the above equation in a more general form so that we can apply Lagrange inversion formalism to obtain $\epsilon(\mathcal{J})$
	\begin{eqnarray}\label{gme}
	\epsilon=\epsilon_0~ \exp\left[\sum_{n=1}^{\infty} \text{a}_n(\mathcal{J}) \epsilon^n\right],
	\end{eqnarray}
	where the coefficients $\text{a}_n$ can be determined from (\ref{gmlne}) and 
	\begin{eqnarray}
	\epsilon_0=4~ \exp\left[-\mathcal{J}-1+B\right],
	\end{eqnarray}
	solves (\ref{gmlne}) to lowest order in $\epsilon$. 
	We write the inverted $\epsilon(\mathcal{J})$-series in terms of Lambert $\mathbf{W}$-function after performing inversion by Lagrange-B$\ddot{\text{u}}$rmann formula  
	\begin{eqnarray}
	\epsilon^2_{\text{leading}}=\frac{2}{\mathcal{J}}\mathbf{W}\left(8\mathcal{J}\exp(-2-2\mathcal{J}+2B)\right).
	\end{eqnarray}
	Finally the dispersion relation for spinning folded string become
	\begin{eqnarray}
	\mathcal{E}-\mathcal{J}=1-B+\frac{\sqrt{2}}{\sqrt{\mathcal{J}}} B \sqrt{\mathbf{W}}-\frac{1}{2\mathcal{J}}\left(\mathbf{W}+\frac{\mathbf{W}^2}{2}\right)+...
	\end{eqnarray}
	The result reduces to (\ref{infgkp}) for $\mathcal{J}=\infty$ and putting $B=0$, it agrees well with the finite size dispersion relation of spinning folded string \cite{Floratos:2013cia}. 
	\section{Conclusions}
	In this paper, we have studied various semiclassical strings which are dual to the long $\mathcal{N}=4$ SYM operators in $R\times S^2$ background in presence of NS-NS flux. The classical dispersion relation for finite size giant magnon carrying large angular momentum and single spike carrying large $p$ has been computed in terms of Lambert $\mathbf{W}-$function. We have also calculated the classical exponential corrections to the  infinite size dispersion relation for GKP string and expressed the result in terms of Lambert $\mathbf{W}-$function. GKP string dispersion relations provide the scaling dimensions of the long $\mathcal{N}=4$ SYM operators Tr$[\phi z^m\phi z^{J-m}]+...$ at strong coupling. The presence of NS-NS B-field does not make any geometrical changes since the equations of motion of strings are independent of B-field. It changes the general form of dispersion relations as it appear in conserved angular momentum. The calculations performed here could be extended in a few directions. First of all, it will be interesting to generalize the finite size corrections to the case of other general rotating strings in AdS with and without the NS-NS field. It will also be interesting to repeat the analysis for rotating and pulsating strings in orbifolded backgrounds. We will come back to some of these issues in future. 
	
	\appendix
	\appendixpageoff
	\section{Appendix}
	In this appendix, we write the results which have been computed by using Mathematica. 
	\subsection*{Giant Magnon:}
	\begin{eqnarray}
	z_{max}=\sin a&=&\sin \frac{p}{2}+\frac{1}{4}\cos^2\frac{p}{2}\left[2\mathcal{J}+(3-2B)\sin\frac{p}{2}\right]\epsilon^2+\frac{1}{4}B\cos\frac{p}{2}\sin p \, \epsilon^3- \nonumber\\
	&-&\frac{3}{64}\cos^2\frac{p}{2}\left[8\mathcal{J}^2\sin \frac{p}{2}-\mathcal{J}\left(12\cos p +B\frac{16}{3}(1-2 \cos p)\right)-5\sin \frac{3p}{2}+ \right.\nonumber\\
	&+&\left.B(4+16\cos p)\sin \frac{p}{2} \right] \epsilon^4+...
	\end{eqnarray}
	We inverted the series (\ref{lne}) for $\epsilon(p,\mathcal{J})$ and then squaring it we have the following expression 
	\begin{eqnarray}
	\epsilon^2&=&16e^{-2L}-128 B e^{-3L}+\left[256\mathcal{J}^2\cot^2\frac{p}{2}+\mathcal{J}\left[64\csc\frac{p}{2}(3\cos p+1)-256B \cos\frac{p}{2}\cot\frac{p}{2}\right]-\right. \nonumber\\
	&-& \left. \vphantom{\sum} 128(1-B)\right]e^{-4L}-B\left[5120\mathcal{J}^2\cot^2\frac{p}{2}+\mathcal{J}\csc\frac{p}{2}(3328\cos p+768)-2048B\right]e^{-5L}+ \nonumber\\
	&+&\left[6144\mathcal{J}^4\cot^4\frac{p}{2}+\mathcal{J}^3\cot^2\frac{p}{2}\csc\frac{p}{2}\left[9728\cos p+512-12288 B\cos^2\frac{p}{2} \right]-\mathcal{J}^2\left[512\csc^2\frac{p}{2}+\right.\right.\nonumber \\
	&+&\left.\left.8448\cos p+6400-B\cot^2\frac{p}{2}\left(7168-11264\cos p\right)\right]+\mathcal{J}\csc\frac{p}{2} \left[
	384 \cos 2p -3264 \cos p -1472-\right.\right.  \nonumber\\
	&-&\left.\left. \vphantom{\sum} B(768\cos 2p-6400 \cos p-4096)\right]+960-2176B \right]e^{-6L}+...
	\end{eqnarray}
	where $L=\mathcal{J}\csc\frac{p}{2}+1-B$.
	Substituting $\epsilon(p,\mathcal{J})$ into corresponding formula, we obtain the finite size dispersion relation for giant magnon
	\begin{eqnarray}
	\mathcal{E}-\mathcal{J}&=&(1-B)\sin\frac{p}{2}+4B\sin \frac{p}{2}e^{-L}-\left[4\sin^3 \frac{p}{2} +4B\cos^2\frac{p}{2}\left(2\mathcal{J}+3\sin \frac{p}{2}\right)\right]e^{-2L}+\nonumber\\
	&+&B\left[8\mathcal{J}^2\csc^3 \frac{p}{2}\sin^2 p+\mathcal{J}(24+40\cos p)+8(3+\cos p)\sin \frac{p}{2}\right]e^{-3L}-\left[8B \mathcal{J}^3\csc^6 \frac{p}{2}\sin^4 p+\right. \nonumber\\
	&+&\left.\mathcal{J}^2\left(8\csc \frac{p}{2}\sin^2 p+80 B\cos\frac{p}{2}\cot\frac{p}{2}(1+3\cos p)\right)-\mathcal{J}\left(12\cos 2p-8\cos p-4-\right. \right. \nonumber\\
	&-&\left.\left.\vphantom{\sum}8B (9\cos 2p+17\cos p+8)\right)+4\left(6\cos p+7\right)\sin^3\frac{p}{2} +B\left(15\sin\frac{5p}{2}+2\sin\frac{3p}{2}+\right.\right.\nonumber\\&+&\left.\left.\vphantom{\sum}3\sin\frac{p}{2}\right)\right]e^{-4L}+...
	\end{eqnarray}
	\subsection*{Spiky string:}
	\begin{eqnarray}
	\mathcal{E}-\frac{p}{2}&=&\frac{q}{2}+B\tan\frac{q}{2}-4B\tan\frac{q}{2}e^{-Q}+4\left[\sin^2\frac{q}{2}\tan\frac{q}{2}+4B\left( (p+q)-\sin q+\left(2+\sec^2\frac{q}{2}\right)\times\right.\right.\nonumber \\ &\times&\left.\left.\tan\frac{q}{2}\right)\right]e^{-2Q}-B\left[16p^2\csc q+4p \sin\frac{q}{2}\sec^3\frac{q}{2}\left(4q\cot^2\frac{q}{2}+3\sin q\right)+\frac{1}{2}\csc\frac{q}{2}\sec^3\frac{q}{2}\times \right. \nonumber \\
	&\times&\left.\vphantom{\sum}\left(10+8q^2-(15-8q^2)\cos q+6\cos 2q-\cos 3q+12q\sin q-6q\sin 2q\right)\right]e^{-3Q}+ \nonumber\\
	&+&\frac{1}{8}\left[128 B p^3 \csc^2\frac{q}{2}+64 p^2\tan\frac{q}{2} \left(1 + 
	B \left(4 + 3\sec^2\frac{q}{2} - 2 \cot^2\frac{q}{2} + 3 q \csc^4\frac{q}{2} \sin q\right)\right) +\right.\nonumber\\
	&+&\left.\vphantom{\sum} p\sec^2\frac{q}{2}\left(16 (3 +(-4 +\cos q)\cos q+ 4 q \sin q)+B \csc^2\frac{q}{2} \sec^2\frac{q}{2} (187 +144 q^2 +\right. \right.\nonumber\\
	&+&\left.\left.\vphantom{\sum} 2(-107 + 96 q^2) \cos q + 
	8 (1 + 6 q^2) \cos 2q+ 22\cos 3q-3\cos 4q - 
	8 q (-13 \sin q + 10 \sin 2q +\right. \right.\nonumber\\
	&+&\left.\left.\vphantom{\sum} 3 \sin 3q))\right)+2 \sec^2\frac{q}{2}\tan\frac{q}{2} \left((-15 +16 q^2)\cos q+4(3+4q^2+ \cos 2q) - \cos 3q -\right.\right.\nonumber \\
	&-&\left.\left.\vphantom{\sum}8q(-3 +\cos q)\sin q\right)+32 B \left(6 q\cos q - 4 q^2\cot\frac{q}{2}+\frac{1}{16}\sec^4\frac{q}{2}\left(120 q - 96 q^3 + 12 \sin q - \right.\right.\right.\nonumber\\
  &-&\left.\left.\left.\vphantom{\sum}30 \sin 2q + 
 8 q (-4 (4 + q^2) \cos q - 7\cos 2q + 
 2 q (4 q\csc^2\frac{q}{2} + 5 \sin q + \sin 2q)) + 4\sin 3q - \right.\right.\right.\nonumber\\
 &-&\left.\left.\left.\vphantom{\sum}
 \sin 4q + 96\tan\frac{q}{2}\right) \right)\right]e^{-4Q}+...
	\end{eqnarray}
	Where we have defined $Q=\frac{1}{2}(p+q)\cot\frac{q}{2}+B\left(1-(p+q)\csc q\right)$.
	\subsection*{GKP string:}
	\begin{eqnarray}
	\mathcal{E}-\mathcal{J}&=&1-B+4B e^{-\mathcal{R}}-4e^{-2\mathcal{R}}-16B (\mathcal{J}-1)e^{-3\mathcal{R}}+4(4\mathcal{J}-1-4 B)e^{-4\mathcal{R}}+\nonumber\\
	&+& B (160 \mathcal{J}^2-176\mathcal{J}+88)e^{-5\mathcal{R}}-(128 \mathcal{J}^2-32\mathcal{J}+32 - B (256 \mathcal{J}-32))e^{-6\mathcal{R}}-\nonumber\\
	&-&B\left(\frac{6272}{3}\mathcal{J}^3-2240\mathcal{J}^2+1696\mathcal{J}-544\right)e^{-7\mathcal{R}}+\left(\frac{4096}{3}\mathcal{J}^3-256\mathcal{J}^2+608\mathcal{J}-84-\right.\nonumber\\
	&-&\left.\vphantom{\sum}B(4096\mathcal{J}^2-512\mathcal{J}+608)\right)e^{-8\mathcal{R}}+ B (31104 \mathcal{J}^4-31104 \mathcal{J}^3 + 29664\mathcal{J}^2- 14672\mathcal{J} +3508)\times\nonumber\\
	&\times&e^{-9\mathcal{R}}-\left(\frac{51200}{3}\mathcal{J}^4-\frac{5120}{3}\mathcal{J}^3+10368\mathcal{J}^2-1760\mathcal{J}+744-B\left(\frac{204800}{3}\mathcal{J}^3-5120\mathcal{J}^2+\right.\right.\nonumber\\
	&+&\left.\left.\vphantom{\sum} 20736\mathcal{J}-1760\right)\right)e^{-10\mathcal{R}}-B \left(     \frac{1}{15} 7496192 \mathcal{J}^5-\frac{1}{3} 1362944 \mathcal{J}^4+\frac{1}{3}1564288 \mathcal{J}^3-309056\mathcal{J}^2+\right.\nonumber\\
	&+&\left.\vphantom{\sum}124080 \mathcal{J}-23344 \right)e^{-11\mathcal{R}}+\left(\frac{1}{5} 1179648 \mathcal{J}^5+ 178176 \mathcal{J}^3 - 25856 \mathcal{J}^2 +26624 \mathcal{J}-2592 -\right.\nonumber\\
	 &-&\left.\vphantom{\sum}B (1179648 \mathcal{J}^4+534528 \mathcal{J}^2- 51712 \mathcal{J}+26624)\right)e^{-12\mathcal{R}}+...
	\end{eqnarray}
Here, we have substituted $\mathcal{R}=1+\mathcal{J}-B$.
	\section{Lagrange Inversion Formalism}
	Lagrange-B$\ddot{\text{u}}$rmann formula is used to get the Taylor series expansion of the inverse function of an analytic function. Let $z$ is a function of $x$ 
	\begin{eqnarray}\label{af}
	z=f(x),
	\end{eqnarray}
	where $f$ is analytic at point $x=x_0$ and $f(x_0)\ne 0$.\\
	Equation (\ref{af}) can be inverted for $x$ and expressed by a power series
	\begin{eqnarray}
	x=g(z)=x_0+ \sum_{n=1}^{\infty}g_n \frac{(z-f(x_0))^n}{n!},
	\end{eqnarray}
	where 
	\begin{eqnarray}
	g_n=\lim\limits_{x\rightarrow x_0}\left[\frac{d^{n-1}}{dx^{n-1}}\left(\frac{x-x_0}{f(x)-f(x_0)}\right)^n\right].
	\end{eqnarray}
	When the function $f$ has the form $f(x)=\frac{x}{y(x)}$, where $y(x)$ is a analytic function with $y(0)\ne 0$ and taking $x_0=0$, the inverse series $g(z)$ is given by
	\begin{eqnarray}
	g(z)=\sum_{n=1}^{\infty}\left(\lim\limits_{x\rightarrow 0}\frac{d^{n-1}}{dx^{n-1}}y(x)^n\right)\frac{z^n}{n!}.
	\end{eqnarray}
	When $y(x)$ has exponential functional form, the invesre series $g(z)$ can be expressed in terms of Bell polynomial.\\
	Partial exponential Bell polynomials are defined as
	\begin{eqnarray}
	\mathcal{B}_{n,k}(\text{a}_1,\text{a}_2,\text{a}_3...\text{a}_{n-k+1})=\sum \frac{n!}{j_1! j_2!...j_{n-k+1}!}\left(\frac{\text{a}_1}{1!}\right)^{j_1}\left(\frac{\text{a}_2}{2!}\right)^{j_2}...\left(\frac{\text{a}_{n-k+1}}{(n-k+1)!}\right)^{j_{n-k+1}}, \nonumber\\
	\end{eqnarray}
	and ordinary Bell polynomials
	\begin{eqnarray}
	\hat{\mathcal{B}}_{n,k}(\text{a}_1,\text{a}_2,\text{a}_3...\text{a}_{n-k+1})=\sum \frac{k!}{j_1! j_2!...j_{n-k+1}!}\text{a}_1^{j_1}\text{a}_2^{j_2}...\text{a}_{n-k+1}^{j_{n-k+1}}, \nonumber\\
	\end{eqnarray}
	where $j'$s are non-negative integers and the sum is taken over all $j$ with simultaneously these two conditions are satisfied
	\begin{eqnarray}
	&& j_1+j_2+j_3+...+j_{n-k+1}=k,\nonumber\\
	&&j_1+2j_2+3j_3+...+(n-k+1)j_{n-k+1}=n.
	\end{eqnarray}
	We can write ordinary Bell polynomials in terms of exponential Bell polynomials by the relation
	\begin{eqnarray}
	\hat{\mathcal{B}}_{n,k}(\text{a}_1,\text{a}_2,\text{a}_3...\text{a}_{n-k+1})=\frac{k!}{n!}\,\mathcal{B}_{n,k}(1!\text{a}_1,2!\text{a}_2,3!\text{a}_3...(n-k+1)!\text{a}_{n-k+1}).
	\end{eqnarray}
	\section{Lambert $\mathbf{W}$-Function}
	Lambert $\mathbf{W}$-function which is also known as omega function satisfy the following relation
	\begin{eqnarray}
	\mathbf{W}(z)e^{\mathbf{W}(z)}=z.
	\end{eqnarray}
	The series expansion of the principal branch of $\mathbf{W}(z)$
	 is given by
	\begin{eqnarray}
	\mathbf{W}(z)=\sum_{n=1}^{\infty} \frac{(-n)^{n-1}}{n!}z^n.
	\end{eqnarray}
	We write few series involving $\mathbf{W}$-function which are useful in our calculation
	\begin{eqnarray}
	\sum_{n=1}^{\infty}(-1)^{n+1}\frac{n^{n}}{n!}z^n=\frac{\mathbf{W}(z)}{1+\mathbf{W}(z)},\nonumber\\
	\sum_{n=1}^{\infty}(-1)^{n+1}\frac{n^{n-2}}{n!}z^n=\mathbf{W}(z)+\frac{\mathbf{W}(z)^2}{2}.
	\end{eqnarray}
	\section{Useful Elliptic Integrals and Jacobi Elliptic Functions }
	In this appandix we collect some relevant formulas we use in this paper. The incomplete elliptic integral of first kind is defined as
	\begin{eqnarray}
	\mathbb{F}(\varphi, m)=\int_0^ \varphi \frac{d \theta}{\sqrt{1-m \sin^2 \theta}},
	\end{eqnarray}
	where the range of modulus m and amplitude $\varphi$ are $0\leq m\leq 1$ and $0 \leq \varphi \leq \frac{\pi}{2}$ respectively.
	We write complete elliptic integral when amplitude $\varphi=\frac{\pi}{2}$,
	\begin{equation}
	\mathbb{K}(m)=\mathbb{F}(\frac{\pi}{2}, m).
	\end{equation}
	
	Similarly the elliptic integral of second and third kind are written as
	\begin{eqnarray}
	\mathbb{E}(\varphi,m)&=&\int_{0}^{\varphi} d\theta\sqrt{1-m \sin^2 \theta},~~~~~~~~ \mathbb{E}(m)=\mathbb{E}(\frac{\pi}{2},m), \nonumber\\
	\boldmath{\Pi}(n,\varphi,m)&=&\int_{0}^{\varphi} d\theta\frac{1}{(1-n \sin^2 \theta)\sqrt{1-m \sin^2 \theta}},~~~~~ \boldmath{\Pi}(n,m)=\boldmath{\Pi}(n,\frac{\pi}{2},m). \nonumber\\
	\end{eqnarray}
	Here are some formulas, we need for our calculation
	\begin{eqnarray}
	\int_{z_{min}}^{z_{max}} dz \frac{1}{\sqrt{(z_{max}^2-z^2)(z^2-z_{min}^2)}}=\frac{1}{z_{max}}\mathbb{K}\left(1-\frac{z_{min}^2}{z_{max}^2}\right), \nonumber \\
	\int_{z_{min}}^{z_{max}} dz \frac{z^2}{\sqrt{(z_{max}^2-z^2)(z^2-z_{min}^2)}}=z_{max}\mathbb{E}\left(1-\frac{z_{min}^2}{z_{max}^2}\right), \nonumber \\
	\int_{z_{min}}^{z_{max}} dz \frac{1}{(1-z^2)\sqrt{(z_{max}^2-z^2)(z^2-z_{min}^2)}}=\frac{1}{z_{max}(1-z_{max}^2)}\boldmath{\Pi}\left(\frac{z_{max}^2-z_{min}^2}{z_{max}^2-1},1-\frac{z_{min}^2}{z_{max}^2}\right),\nonumber \\
	\mathbb{F}(\varphi,m)=\frac{1}{\sqrt{m}}\mathbb{F}(\varphi_1,\frac{1}{m})~~~~~\text{where}~ \varphi_1=\arcsin(\sqrt{m}\sin\varphi),\nonumber \\
	\boldmath{\Pi}(n,m)=\frac{1}{(1-n)\mathbb{K}(1-m)}\left[\frac{\pi}{2}\sqrt{\frac{n(n-1)}{m-n}}\mathbb{F}\left(\arcsin\sqrt{\frac{n}{n-m}},1-m\right)- \right. \nonumber \\
	-\left.\mathbb{K}(m)\left[(n-1)\mathbb{K}(1-m)-n\boldmath{\Pi}\left(\frac{1-m}{1-n},1-m\right)\right]\right].\nonumber\\
	\end{eqnarray}

\providecommand{\href}[2]{#2}\begingroup\raggedright\endgroup
\end{document}